\documentclass{iau}

\usepackage{amsmath}
\usepackage{graphicx}
\usepackage{multirow}
\usepackage{hyperref}

\begin{document}

\title[Panel Discussion: Machine Learning Practical Problem Solving] 
{Panel Discussion: Practical Problem Solving for Machine Learning}

\author[Cabrera et al.]   
{Guillermo~Cabrera$^{1*}$, Sungwook~E.~Hong$^{2*}$, Lilianne~Nakazono$^{3*}$, David~Parkinson$^{2*}$, \and 
Yuan-Sen~Ting$^{4,5*}$}

\affiliation{$^1$ 
Department of Computer Science, University of Concepci\'on, Chile\\
email: {\tt guillecabrera@inf.udec.cl}\\[\affilskip]
$^2$Center for Theoretical Astronomy, Korea Astronomy and Space Science Institute, \\
776 Daedeok-daero, Yuseong-gu, Daejeon 34055, Republic of Korea \\
email: {\tt swhong@kasi.re.kr, davidparkinson@kasi.re.kr} \\[\affilskip]
$^3$ Instituto de Astronomia, Geofísica e Ciências Atmosf\'ericas da U. de S\~ao Paulo\\ Cidade Universit\'aria, 05508-900 São Paulo, SP, Brazil\\ email: {\tt lilianne.nakazono@usp.br} \\[\affilskip]
$^4$ Research School of Astronomy \& Astrophysics, Australian National University,\\ Cotter Road, Weston, ACT 2611, Australia \\email: {\tt yuan-sen.ting@anu.edu.au}\\[\affilskip]
$^5$ School of Computing, Australian National University, Acton ACT 2601, Australia\\[\affilskip]
$^*$Equal contribution
}

\maketitle

\pubyear{2022}
\volume{368}  
\pagerange{119--126}
\setcounter{page}{1}
\jname{Machine Learning in Astronomy:
Possibilities and Pitfalls}
\editors{Ashish Mahabel, Christopher Fluke \& Jess McIver, eds.}

\begin{abstract}
Machine Learning is a powerful tool for astrophysicists, which has already had significant uptake in the community. But there remain some barriers to entry, relating to proper understanding, the difficulty of interpretability, and the lack of cohesive training. In this discussion session we addressed some of these questions, and suggest how the field  may move forward.
\end{abstract}


\firstsection 
\section{Introduction}

The practical utility of machine learning (ML) methods to astrophysical and astronomical problems has already been demonstrated for a multitude of different problems. Yet for many research astronomers, the mechanics of the ML toolbox remains somewhat mysterious, which leads some to use the methods without full understanding, and others to shun ML entirely. The path to adopting ML in astrophysics should be made clearer and easier, but without risking an over-simplification. In this discussion session we address some of the issues of machine learning interpretability, accessibility, and education.

\section{Discussion Questions}

\subsection{What is Interpretability in Astronomical ML Problems?}
\label{sec:interpretability}
\discuss{Sungwook}{We astronomers are automatically astrophysicists. We want to \emph{understand} the nature behind astronomical data, not just handle the data. But most ML techniques that guarantee high predictability are very sophisticated, and so become black boxes for most users.
This means that, ironically, we now need to put additional efforts into \emph{understanding} the very tool that we introduced to \emph{understand} the nature. That is probably one of the key factors why some notable astronomers are disinclined towards applying MLs to astronomy.

While many so-called interpretable MLs have been suggested in the ML community, I believe that we need to discuss with classical, old-fashioned astronomers to define the \emph{interpretability} in astronomical ML. Their expertise is \emph{interpreting} the astronomical data and theories --- their interpretations may not always be valid, but they have a very high probability of being valid. Therefore, if our definition and application of interpretability do not match with their intuition at all, it will not be accepted in the general astronomical society, no matter how fancy it seems to be.}

\discuss{Yuan-Sen}{While interpretability is important for physical sciences, it also comes with many flavours and can vary drastically depending on the context. Take the example of N-body simulations. Complex systems with many ``bodies" are fundamentally top-down processes. In such cases, it would be hard to argue that we could achieve analytic (bottom-up) interpretability. Despite that, I would argue that in many such systems, we can still understand the collective behaviour intuitively without chasing for the illusive bottom-up interpretability. I am often concerned that, in astronomy, we seem to hold a higher standard in terms of interpretability for machine learning than for, say, N-body simulations. I often think of deep learning as more like statistical mechanics. While it might not be possible to disentangle everything down to individual neurons, the collective behaviour of neural networks can be understood as intuitive. Its capability traces back to all the inductive biases rooted in various physical symmetries. And for that, I feel more optimistic about the interpretability issue. I would argue that, at the higher level, we have a decent mathematical understanding of deep learning on its collective or emergent behaviour. And this mathematical foundation of machine learning is bound to be a growing field in the next few years.}

\discuss{Lilianne}{
There is a growing field of research in ML called Explainable AI (XAI) and one can find many systematic reviews in the literature. I particularly recommend start by reading \cite{gilpin2018} 
They state:  
\quote{“The goal of interpretability is to describe the internals of a system in a way that is understandable to humans. […] 
The challenge facing explainable AI is in creating explanations that are both complete and interpretable: it is difficult to achieve interpretability and completeness simultaneously. The most accurate explanations are not easily interpretable to people; and conversely the most interpretable descriptions often do not provide predictive power.”}

In simple terms, an explanation is an answer to a “why” question, i.e. “why a model predicts in a certain way?”. The big “why” questions of the Universe are what inspire us as astronomers. Ideally, our goal should be to try to understand the relations between  e.g. the input-output of a ML model, while also connecting them with theoretical physics. This is not an easy task, especially with the interpretability-completeness trade-off.}

\subsection{Modelling versus Learning}
\label{sec:modelling_vs_learning}

\discuss{Yuan-Sen}{A somewhat misconception about deep learning is that it is a total black box. On the one hand, as discussed above, the success of most, if not all, deep learning architectures can be traced back to their inductive biases and can be understood via their underlying physical symmetries. In this sense, deep learning itself is a type of modelling. On the other hand, any fitting 
even linear regression, is a type of learning. Modelling any physical system is always a delicate balancing act between modelling and learning and is never one or the other. As we impose fewer and fewer inductive biases (learning toward the learning side), as the solution space becomes larger, it will lead to a system that is more ``training-data-hungry" and vice versa.

Viewed through this lens, often, the decider lies in understanding (a) how much knowledge we have on the fundamentals that drive the system and (b) how much data we have. If we have a profound understanding of the system (e.g. the exact equations that describe the system), modelling is favoured because it is less data-hungry. But quite often, we only know the symmetries (or even just approximate symmetries) that the system follows. In this case, perhaps the best we could do is to have a deep learning architecture that satisfies the symmetries. I would argue that this is not a more morally inferior position because choosing such modelling (i.e., the deep learning architecture) reflects our best understanding of the physical systems (the symmetries the systems have). What is more dangerous, however, is a blind application of any deep learning architecture without understanding the inductive biases that the architecture entails. In this case, the deep learning method will starve for data while at the same time subjecting us to the wrong solution space.}

\subsection{Classical Machine Learning versus Deep Learning. When and why.}

\discuss{Lilianne}{
Following the discussion about interpretability-completeness tradeoff in \ref{sec:interpretability}, 
if one is trying to make interpretations from the data at some level, deep learning will be the last thing one would try out. Moreover, not all data and not all problem require complex solutions such as deep learning, so do not blindly go for deep learning (or even machine learning) without clearly understanding your problem. 
Now suppose one is not interested in interpretation but only cares about achieving accurate predictions. I would go for classical machine learning instead of deep learning in those scenarios: (1) when one lacks resources (e.g. GPU, cloud access), as deep learning requires more computational power by its nature; (2) when one has small data sets, as deep learning can easily overfit on them; (3) if one has no previous experience with deep learning or machine learning whatsoever (especially when time is tight!), as building a deep learning architecture can be very tricky, while classical machine learning are often more intuitive and easier to implement. 
Finally, in my point of view, deciding between deep learning and machine learning is also about time and resource management, balancing them with the gains that a deep learning model would provide. For example, if I have to spend 5x more time to train a deep learning model and to put it in production to only improve 0.01\% in accuracy, it is probably not worth it. }

\discuss{Yuan-Sen}{The key distinction between deep learning (i.e., neural network-based) and classical machine learning (regression, PCA, etc.) is that, while not fully understood, deep learning seems to evade the curse of dimensionality. And this is truly profound. As discussed Sec.~\ref{sec:modelling_vs_learning}, a critical goal of many physical modelling is to admit a solution space that is as generic as possible, yet the dimension does not curse it. Briefly, as the dimension increases, it does not require an exponentially large training data set to bound the fitting error. As we have grown to have a much better understanding of the mathematical foundation of deep learning, what has become more apparent is that the size of the solution space and the curse of dimensionality are not mutually exclusive. The fortuitous empirical success of neural networks has demonstrated that, with proper inductive biases, it is possible to admit a valuable ample solution space that is trainable with finite training data. Being able to perform a high-dimensional regression (supervised learning) or describe a high-dimensional probability density function (unsupervised learning) is a game changer that opens up many new possibilities for physical sciences, as most phenomenology in physical sciences is of inherently high dimension.}

\subsection{Tailored Loss Functions for Specific Problems}

\discuss{Sungwook}{As one who prefers applying DL to regression problems, most astronomical DL works seem to focus on enhancing architectures.
However, once the architecture is complex enough, it is hard to achieve significant improvement by fine-tuning architecture or optimization methods.

Instead, I believe we need to focus more onto designing specific loss functions (and/or the normalization method of input and output dataset) optimized for the given problem.
For example, suppose that one reconstructed the matter density using the mean-squared error (MSE) loss function.
Then the machine will try to match the truth and prediction values at each position, but it does not guarantee that the reconstructed density field will have a similar large-scale feature to the ground truth.
Some loss functions contain terms about large-scale features (\cite[Gatys, Ecker \& Bethge 2016]{gatys2016}, \cite[Liu \etal\ 2018]{liu2018}), but they do not guarantee to match the specific large-scale features in astronomical contents, such as power spectra or $N$-point correlation functions.

Another example is the intrinsic bias in the truth sample distribution.
For example, in most region, the 3D dark matter density and the projected 2D dark matter density are around average or somewhat underdense (\cite[Hong \etal\ 2021a]{hong2021a}, \cite[Hong \etal\ 2021b]{hong2021b}).
On the other hand, most information that astronomers want to get come from highly overdense regions, but using usual MSE-like loss functions does not guarantee excellent prediction at such overdense regions.
To overcome this, one should manually emphasize important but less probable regions, such as by normalizing input or output density, adding customized weights to the loss function, etc (\cite[Lin \etal\ 2017]{lin2017}).
In conclusion, I believe astronomers need to focus more on finding loss functions optimized for what they want to match, instead of just adopting well-known functions in the ML community. }

\discuss{Yuan-Sen}{Besides embedding inductive biases (physical symmetries) through the deep learning architecture, another way that has proven effective is to include our understanding of the systems, e.g., equations that the systems satisfy, directly in the loss function. Since we often have a lot more prior knowledge in physical sciences (e.g. PDE that the system follows), and some of these cannot easily materialise at the architecture level, tailoring the loss function to reflect this prior knowledge plays an exceedingly critical role in AI x Science.}

\subsection{What is Necessary for Enhancing Astronomical ML Community?}


 \discuss{Yuan-Sen}{To advance AI x Science, we need to cultivate a community that has a deep appreciation of both domains instead of treating machine learning as a tool that can be applied to astronomy. Deep learning (as opposed to classical machine learning) is a drastic paradigm shift in our understanding of mathematics (e.g., the curse of dimensionality). If the AI x Science community could act collectively to ensure other astronomers also see the true value and excitement in deep learning instead of just portraying deep learning as a tool, I believe it would lead to a positive perception change in the field. This resembles what the computational astronomy (hydrodynamical simulations) community has gone through in many ways. Yes, simulation techniques are useful for science, but at the same time, they are also profound, interesting mathematical objects. Realising the latter has sustained and grown the computational astronomy community drastically, attracting people interested in such mathematics beyond treating it as just a tool.}

\discuss{Lilianne}{Interdisciplinary is the keyword.  Astronomical data sets are an immense (yet underutilised) playground for many theoretical researchers in Computer Science, Statistics, and other STEM areas. Many of the problems we face with astronomical data (e.g. dealing with measurement errors, sample bias, etc) can serve as motivation for developing new techniques to address real-world problems. Therefore, if we want to enhance astronomical ML community, we must strengthen collaborations with other STEM fields. Building bridges between different fields are not easy, requiring an effort from all parts involved. And, of course, it is also important to encourage astronomy students who lean towards interdisciplinary.  
Being a graduate student myself and having my own concerns about my future in academia, I bring here some questions to think about: Is the academic system ready to really embrace cross-disciplinary people? Are evaluation metrics for academic productivity and tenure-tracked positions biased towards another profile? What cultural/systematic changes do we need in our departments, to start with?}

\subsection{How to effectively train students in ML?}

\discuss{Sungwook}{For me, the best way to learn astronomical ML is to (1) apply well-established ML techniques to astronomical problems, not the tutorial problems famous in the ML community, (2) see how they \emph{fail}, and (3) find out how to improve it.
When I first learned CNN, I was surprised at how easy its Python manifest is and how well such simple code can predict the number so well in the MNIST dataset.
When I applied the same code to my problem later, I was shocked that it just did not work.
But, this made me start thinking about how the dataset, architecture, loss function, and optimization method should be treated for the specific problem.
Without such experience, I might still think of ML as either almighty black magic or obsolete, and probably all my ML-related astronomical works would be meaningless.}

\discuss{Yuan-Sen}{I fear that the way that we are currently training students in astronomy might not be sufficient. To cultivate a well-versed community in both domains, we need a more flexible education system that will allow students to cross-pollinate between the two fields. This might mean we might want to consider other options in the PhD program, such as a joint PhD in astronomy and computer science that require students to publish not only in top astronomical journals but also in top machine learning conferences. I have seen such a movement at the Master level, which is certainly moving in the right direction.}

\discuss{Lilianne}{I believe that building the mathematical basis necessary for understanding what is behind ML implementations is more important than teaching ML code recipes. Teaching the foundations gives them freedom to learn (and produce!) more specific knowledge with quality, which could be ML or other things. I can give an example from my personal experience: my first contact with deep learning was through hands-on workshops. The analogies, the schema of neurons and layers, the frameworks, those were all well installed in my brain. However, I remember that I only truly understood deep learning years later after attending a 1-hour seminar where the speaker decided to show the mathematics behind it, with all the demonstrations. That relatively quick moment of realization was only possible because I spent a great amount of time attending fundamental courses. 
Looking at this question by other prism, I would actually give more emphasis on our failure in not cultivating stronger collaborations with machine learning researchers. Science is collaborative and we can make the best of it by working together, combining our expertise. One thing that can be harder than learning how to fit a ML model in Python is learning how to communicate with researchers from other STEM fields, as we all use different jargon.}

\subsection{Ethics of AI in Astronomy}

\discuss{Sungwook}{I agree that students need to learn the ethics of AI, mainly because students can develop their careers outside astronomy.
However, I doubt if faculties in astronomy departments are best for teaching the ethics of AI.
The ethics of AI rapidly changes over time mainly due to the various social issues such as human rights, which most astronomers do not (need to) care about much.
Instead, since all fields will require AI as mandatory in the near future, I think a general AI ethics course should be open at the university level, not the department level.}

\discuss{Lilianne}{When we talk about Ethics of AI we are talking about the impact of AI and data-driven solutions on people’s lives. From \cite{aiethics},

\quote{“
AI ethics [...] is about the ethical challenges posed by current and near-future AI and its impact of our societies and vulnerable democracies. AI ethics is about the lives of people and it is about policy.”}

A classical example, based on the famous trolley problem, is the self-driving car’s choice when faced with multiple, potentially lethal choices. 
Luckily for us astronomers, the worst we can do with AI in Astronomy is telling a galaxy “you are a star!”. No celestial object’s “lives” will ever be affected by our AI models. Actually, we have a bigger elephant in the room: everyone is susceptible to being harmed by the unethical behaviours of colleagues or supervisors.  Unfortunately, the odds are higher when you are from a minority group. I sadly do not have enough fingers to count the number of astronomy students that I know that suffered from moral and/or sexual harassment in academia. Science is done by people, therefore ethics is involved in everything that we produce, not only where AI is involved. A better workplace in academia is something that we must strive for, thus ethics is something that we all must take lessons on.}

\discuss{Yuan-Sen}{While deep learning has helped astronomy tremendously in the last few years, what perhaps has not been emphasised enough is that the astronomy dataset is the perfect dataset for the deep learning community to latch on to. On the one hand, underlining our data is the myriad of physical symmetries bound to challenge and advance our mathematical understanding of deep learning. On the other hand, in almost all cases, the astronomy dataset is free from many ethical concerns faced by the deep learning community. Making this point across will undoubtedly create a win-win situation for both communities.}



\begin{thebibliography}{}

\bibitem[Coeckelbergh (2020)]{aiethics}{Coeckelbergh, M. 2020,  \textit{AI Ethics}, The MIT Press, ISBN=9780262357067}

\bibitem[Gatys, Ecker \& Bethge (2016)]{gatys2016} 
{Gatys, L.~A., Ecker, A.~S., \& Bethge, M.} 2016, \textit{in 16th Annual Meeting of the Vision Sciences Society (VSS 2016)}, p. 326, Scholar One, Inc.

\bibitem[Gilpin et al (2018)]{gilpin2018}{Gilpin, L.~ H., Bau, D., Yuan, B.~Z., Bajwa, A., Specter, M. and Kagal, L. 2018, \textit{The 5th IEEE International Conference on Data Science and Advanced Analytics (DSAA 2018)}
}

\bibitem[Hong \etal\ (2021a)]{hong2021a} 
{Hong, S.~E., Jeong, D., Hwang, H.~S., \& Kim, J.} 2021, \textit{Astrophysical Journal}, 913, 76

\bibitem[Hong \etal\ (2021b)]{hong2021b} 
{Hong, S.~E., Park, S., Jee, M.~J., Bak, D., \& Cha, S.} 2021, \textit{Astrophysical Journal}, 923, 266

\bibitem[Lin \etal\ (2017)]{lin2017}
{Lin, T.-Y., Goyal, P., Girshick, R., He, K., \& Doll\'{a}r, P.} 2017, \textit{in IEEE
International Conference on Computer Vision (ICCV)}, ed. K. Ikeuchi,
G. Medioni, \& M. Pelilo (New York: IEEE), 2999

\bibitem[Liu \etal\ (2018)]{liu2018}
{Liu, G., Reda, F.~A., Shih, K.~J., Wang, T.~C., Tao, A., \& Catanzaro, B.} 2018, \textit{European Conference on Computer Vision}, pp. 89-105, Springer, Cham


\end{thebibliography}
\end{document}